\documentclass[12pt]{iopart}
\usepackage{graphicx}
\usepackage{epstopdf}
\usepackage{iopams} 
\usepackage{multirow} 

\let\a=\alpha \let\b=\beta   
   
\let\l=\lambda \let\m=\mu \let\n=\nu   
\let\s=\sigma

\def\nn{\nonumber}

\def\be{\begin{equation}}
\def\ee{\end{equation}}
\def\bea{\begin{eqnarray}}
\def\eea{\end{eqnarray}}
\def\ba{\begin{array}}
\def\ea{\end{array}}

\def\td{\tilde}

\begin{document}

\title{Collision-dominated spin transport in graphene and Fermi liquids}

\author{Markus M\"uller$^{1,3}$ and Hai Chau Nguyen$^{1,2}$}
\address{$^{1}$International Center for Theoretical Physics, Strada Costiera 11, 34014 Trieste, Italy}
\address{$^{2}$Institute for Theoretical Physics, University of Cologne, 50937 Cologne, Germany}
\ead{$^{3}$markusm@ictp.it}

\begin{abstract}
In a clean Fermi liquid, due to spin up/spin down symmetry, the dc spin current driven by a magnetic field gradient is finite even in the absence of impurities. Hence, the spin conductivity $\sigma_{s}$ assumes a well-defined collision-dominated value in the disorder-free limit, providing a direct measure for the inverse strength of electron-electron interactions. 
In neutral graphene, with Fermi energy at the Dirac point,  the Coulomb interactions remain unusually strong, such that the inelastic scattering rate comes close to a conjectured upper bound  $\tau_{\rm inel}^{-1} \lesssim k_BT/\hbar$, similarly as in strongly coupled quantum critical systems.
The strong scattering is reflected by a minimum of the spin conductivity at the Dirac point, where it reaches $\sigma_{s} = \frac{0.121}{\alpha^2}\frac{\mu_s^2}{\hbar}$ at weak Coulomb coupling $\alpha$, $\mu_s\approx \mu_B$ being the magnetic moment of the electronic spins. Up to the replacement of quantum units, $e^2/\hbar \to \mu_s^2/\hbar$, this result equals the collision-dominated  {\em electrical} conductivity obtained previously. This accidental symmetry is, however, broken to higher orders in the interaction strength.
For gated graphene, and 2d metals in general, we show that the transport time is parametrically smaller than the collision time. We exploit this to compute the collision-limited $\s_s$ analytically 
as
$\sigma_{s}= \frac{1}{C} \left(\frac{\mu}{T}\right)^2  \frac{\mu_s^2}{\hbar}$ with $C=4\pi^2\alpha^2\left[\frac{2}{3}\ln(1/2\alpha)-1\right]$ for weak Coulomb coupling $\alpha$.
\end{abstract}


\maketitle


\section{Graphene - a strongly coupled, relativistic electron-hole plasma}

Graphene is a monolayer of graphite, which forms a zero-gap semiconductor whose low energy quasiparticles obey the massless Dirac equation~\cite{Semenoff84,Haldane88,rmp}. At charge neutrality, the Fermi surface reduces to two inequivalent Fermi points, forming a non-analyticity in the density of states, which can be viewed as a very simple quantum critical point~\cite{joerg}. However, on top of that, as a consequence of the vanishing density of states and the linear dispersion of the 2d quasiparticles Coulomb interactions remain unusually strong. They are only marginally irrelevant under renormalization, flowing only logarithmically to zero with decreasing temperature $T$~\cite{guinea}. Moreover, screening is suppressed so that the long range character of the Coulomb interactions remains mostly intact, except for screening due to thermally excited carriers. 
This strong interaction is reflected, e.g., in the inelastic transport scattering rate being proportional to
$N \alpha^2 T$, where $N=4$ is the number of fermionic species, and we use units with $k_B =1$. Here, $\alpha=e^2/\kappa \hbar v_F$ is the dimensionless "fine structure constant" characterizing the strength of Coulomb interactions, where $\kappa$ is the average dielectric constant of the adjacent medium and $v_F$ is the Fermi velocity of the linearly dispersing quasiparticles.
The large scattering rate in graphene comes close to saturating a kind of Heisenberg uncertainty principle for quasiparticles~\cite{subirbook}. This latter is the conjecture that the scattering rate can never  significantly exceed the thermal energy scale, $\hbar \tau^{-1}_{\rm inel}\lesssim  T$. Upon approaching this limit by dialing up $\alpha\to O(1)$, one indeed expects to drive a quantum phase transition towards an insulator~\cite{Laehde, Herbut, Son} with completely different low energy excitations.

Due to the strong marginal interactions, the neutrality point of graphene exhibits a transport phenomenology very similar to quantum critical points in more complex, strongly coupled materials~\cite{joerg,qcgraphene,ssqhe,Damle, Bhaseen}. What makes graphene particularly attractive in this context is the fact that it is probably the simplest condensed matter system to possess the ingredients for this strong coupling phenomenology, being at the same time rather easy to produce experimentally at a high level of purity~\cite{CleanSuspendedBolotin,andrei}. Indeed, a spectacular experimental proof for the non-negligible Coulomb interactions in clean graphene is the recent observation of the fractional quantum Hall effect~\cite{FQHEAndrei,FQHEBolotin,AbaninAndrei09}.

The eventual logarithmic flow of the Coulomb coupling constant $\a$ toward zero justifies a perturbative analysis of the transport properties at low enough temperatures. Upon extrapolation to moderate coupling this captures the gist of the transport phenomenology, even though the results should not be trusted at a quantitative level beyond the weak coupling regime.
In the "quantum critical window", i.e., at small chemical potential of the carriers, $|\mu| < T$, the quasiparticles of graphene form an interacting "hot" (non-degenerate) electron-hole plasma with rather unusual transport properties, such as an anomalously low viscosity and a concomitant tendency toward turbulent electronic current flow in mesoscopic samples.~\cite{viscosity} Those are features one should equally well expect in strongly coupled quantum critical matter, if it is sufficiently clean. Another interesting consequence of the strong scattering in graphene is the relativistic hydrodynamic transport which emerges in a regime of moderately low frequencies below the inelastic collision rate.~\cite{MHDgraphene,cyclotron}  

At finite carrier density, a simple estimate of the inelastic scattering rate in random phase and Born approximation suggests that
\bea
\hbar \tau^{-1}_{\rm inel} \sim {\rm max} (T,|\mu|) \frac{\alpha^2}{(1+\alpha|\mu|/T)^2},
\eea
where $\alpha\approx \alpha(\epsilon) = \alpha_\Lambda/[1+\alpha_\Lambda/4 \ln (\Lambda/\epsilon)]$ denotes the renormalized strength of Coulomb interactions $\alpha$ at the relevant energy scale $\epsilon={\rm{max}}[\mu,T]$, whereby $\Lambda$ is a UV cutoff~\cite{guinea}. (In the sequel we simply write $\alpha$ with the understanding that this renormalized value should be used.)
At finite $\mu$, the scattering rate decreases rather quickly with $T$, following the familiar law $T^2/|\mu|$, which we will show to be merely logarithmically dependent on the interaction strength $\alpha$.

As was first pointed out in the context of the superfluid-insulator quantum phase transition~\cite{Damle} the particle-hole symmetric point $\mu=0$ exhibits a finite collision-dominated conductivity, even in the absence of impurities. Indeed, the application of an external electrical field induces counter propagating particle and hole currents, and thus no net momentum. The latter is usually the source of infinite current response unless the momentum decays due to impurities, Umklapp scattering being negligible at the temperatures we have in mind. However, in neutral graphene one finds a disorder-independent conductivity which is solely due to electron-hole friction. Scaling arguments~\cite{Ryzhii,qcgraphene} based on the Drude formula using  the thermal density of carriers $n_{\rm th}\sim (T/\hbar v_F)^2$, the above discussed inelastic scattering rate, and a $T$-dependent effective mass $m_{\rm eff}\sim T/v_F^2$ suggest a conductivity which is nearly constant, apart from a weak logarithmic growth due to the renormalization of $\alpha$,
\bea
\sigma_e(\mu=0)\sim \frac{e^2n_{\rm th} \tau_{\rm inel}}{m_{\rm eff}} = \frac{\kappa}{\alpha(T)^2} \frac{e^2}{\hbar}.
\eea
This is indeed confirmed by a microscopic calculation based on the semiclassical Boltzmann equation, which becomes asymptotically exact for $T\ll \Lambda$ where $\alpha\ll 1$, yielding the prefactor $\kappa=0.121$~\cite{qcgraphene}. 
\footnote{This weak coupling approach could be extended to include screening effects along the lines of the recent work~\cite{Schuett10}, by considering the simultaneous limit $\alpha\to 0$ and $N\to \infty$, keeping $\a N$ constant. In this limit the random phase approximation becomes exact and could be incorporated in the kinetic equation. The above calculation, as well as the one in the present paper,  correspond to the limit $\a N\ll 1$ of such an approach. On the other hand, for $\a N\gg 1$  one will find a scattering rate $\hbar \tau_{\rm inel}^{-1}\sim T/N$ and a conductivity $\sigma_e \sim N^2$, while remaining in a regime where the perturbative weak coupling approach is still justified.}  


A similar phenomenon arises when we consider spin transport in the presence of magnetic field gradients. A finite, collision-limited spin conductivity $\s_s$ or spin diffusion constant $D_s$  arises whenever the symmetry between spin up and spin down is not broken by a background field. Unlike the electrical conductivity, which diverges at finite gating in the disorder free-case, spin diffusion may serve as an interesting and direct measure of electron-electron interactions in clean metals in general. In this work we analyze spin transport in the linear response regime as a function of gating. This is motivated in particular by recent measurements of the gating dependence of spin diffusion in graphene~\cite{vanWees09,vanWees10}, which found the spin diffusion constant to be roughly proportional to the charge diffusion constant. In the relatively disordered samples under study, this proportionality was attributed to impurities which presumably affect the mean free path of both diffusion processes in a similar way. However, this proportionality is rather unlikely to survive in clean, suspended graphene where interactions are expected to become the dominant source of scattering. Clearly the simple proportionality cannot hold at finite gating where $\s_e$ remains sensitive to impurities, while $\s_s$ is entirely collision dominated in the weak disorder limit. As we will see, also in the ungated case, at the Dirac point, the comparison of the two diffusion constants may provide an interesting probe for strong coupling effects.  
    
\begin{figure}
\begin{center}
\includegraphics[width=8.5cm]{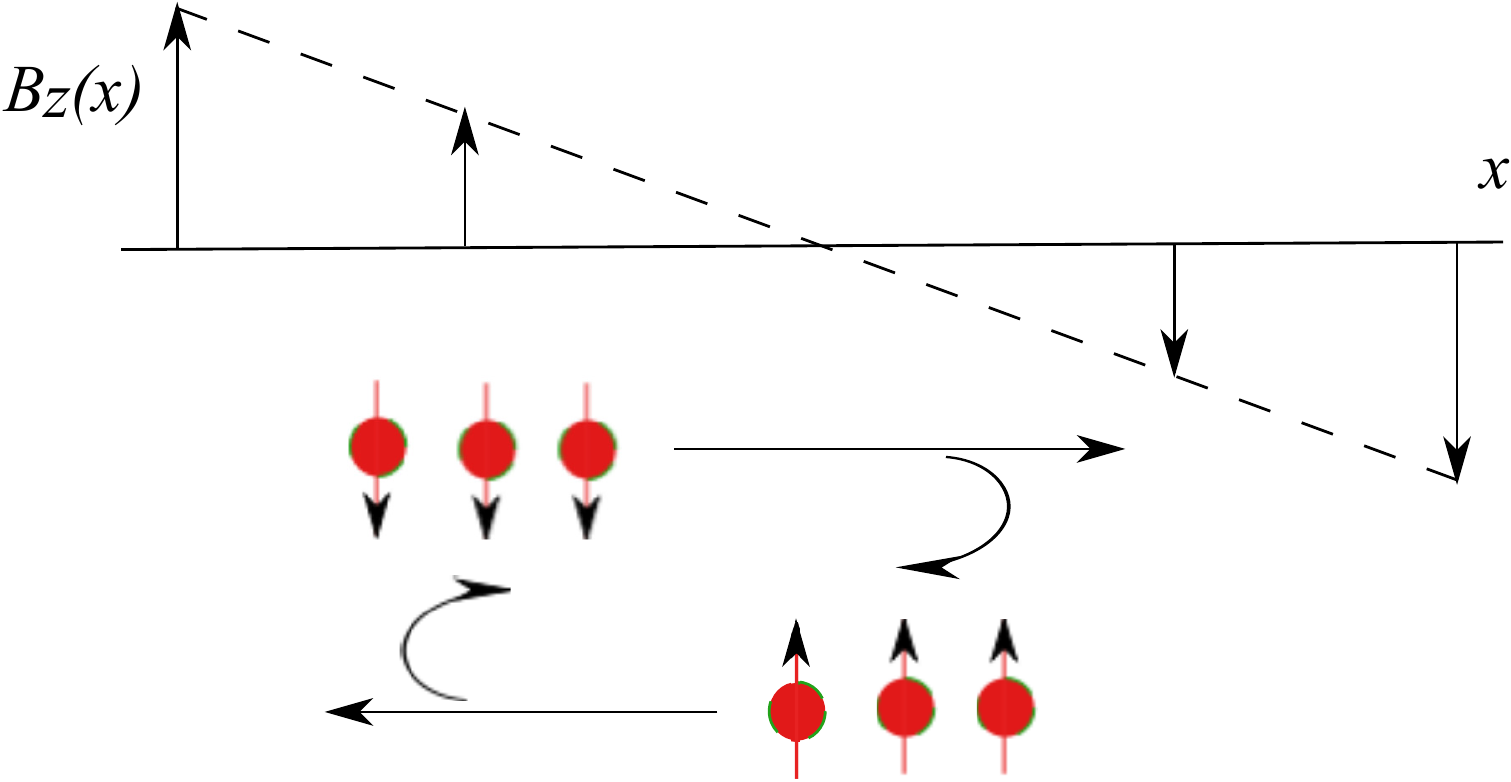}
 \caption{Finite collision-dominated spin transport. In the presence of a magnetic field gradient spin down carriers are accelerated opposite to spin up carriers, creating a counterflow with vanishing total momentum.  Interactions among the carriers lead to friction between the counter propagating up and down currents, which degrades the spin current and leads to a purely interaction-limited spin conductivity. 
}
\label{fig:spinconductivity}
\end{center}
\end{figure}

\section{Model}
At low energies, graphene is described by massless Dirac quasiparticles obeying~\cite{rmp}
\begin{equation}
H_0 = v_F \sum_{\n,\s=\pm} \int \mathrm{d}^2r \, \psi_{\n\s}^{\dagger} \vec{\tau}\cdot \vec{p}\, \psi_{\n\s} 
=\sum_{\n,\s,\l=\pm}\int \frac{\mathrm{d}^2k}{(2 \pi)^2} \lambda \hbar v_F k \gamma_{\n\s \lambda \vec{k} }^{\dagger} \gamma_{\n\s \lambda \vec{k}} ,
\end{equation}
where $v_F$ is the Fermi velocity and $\tau_{x,y}$ are the Pauli matrices in sublattice space. The indices $\{\n,\sigma,\l\}$, label the valley ($\n=\pm$ for $K$ or $K'$ point, respectively),  the electron spin, and  the valence or conduction sub-band, respectively. The quasiparticle operators $\gamma_{\n\s\l\vec{k}}$ diagonalize the Hamiltonian. 

In this basis, electron-electron interactions are described by~\cite{qcgraphene}
\begin{equation}
H_1 =\sum_{ij} \sum_{\lambda_1 \lambda_2 \lambda_3 \lambda_4} \int \frac{\mathrm{d}^2 k_1}{(2 \pi)^2}\frac{\mathrm{d}^2 k_2}{(2 \pi)^2}\frac{\mathrm{d}^2q}{(2 \pi)^2} T_{\lambda_1 \lambda_2 \lambda_3 \lambda_4} (\vec{k}_1,\vec{k}_2,\vec{q}) \gamma_{i \lambda_4 \vec{k}_4}^{\dagger}\gamma_{j \lambda_3 \vec{k}_3}^{\dagger}\gamma_{j \lambda_2 \vec{k}_2 } \gamma_{i \lambda_1 \vec{k}_1},
\end{equation}
where $i,j$ stand for index pairs $(\nu,\s)$, and
\begin{equation}
\label{Ts}
T_{\lambda_1 \lambda_2 \lambda_3 \lambda_4} (\vec{k}_1,\vec{k}_2,\vec{q}) = \frac{V(q)}{8} \left[ 1 + \lambda_4 \lambda_1 \frac{K_4 ^{\ast}}{|K_4|} \frac{K_1}{|K_1|} \right]\left[ 1 + \lambda_3 \lambda_2 \frac{K_3 ^{\ast}}{|K_3|} \frac{K_2}{|K_2|} \right],
\end{equation}
$V(q)$ being the Fourier transform of the two particle interaction.
Here and in the following we will use the notations $\vec{k}_3 = \vec{k}_2 -\vec{q} $ and $\vec{k}_4 = \vec{k}_1 +\vec{q}$, as well as $K_i = k^x_i+ik^y_i$. 

Below we will mostly consider Coulomb interactions,
\begin{equation}
V^0_{\rm Cb}(q) = \frac{2\pi e^2}{q}=\frac{2 \pi \hbar v_F \alpha}{q}, 
\end{equation}
where the "fine structure constant" $\alpha$ has a slow logarithmic flow under RG transformation as mentioned above. We will be mostly concerned with a weak coupling analysis to leading order in $\alpha$. Close to the Dirac point, $\mu\lesssim T$, the inclusion of screening yields only subdominant corrections and has thus not been included in earlier studies on collision-limited transport in that regime~\cite{qcgraphene,kashuba}.
However, in the gated case, $\mu\gg T$, the screening of long range interactions limits the scattering rate at small momentum transfers. Since our results do not depend crucially on the form of this cut-off, we content ourselves with a static (Thomas-Fermi) screening approximation of the Coulomb interaction 
\begin{equation}
\label{RPA}
V(q)=V_{\rm Cb}^{\rm TF}(q) = \frac{2 \pi \alpha}{q+q_0}.
\end{equation}
Here $q_0 = 8 \alpha T \ln[2\cosh(\mu/2T)]$ is the Thomas-Fermi wave vector, which tends to $q_0 = 4 \alpha \mu$ in the limit $\mu\gg T$. A detailed analysis of screening effects at the Dirac point, including dynamical effects, has been given in the recent work~\cite{Schuett10}. 
\footnote{We note that the approximate analysis of collision-limited transport in Ref.~\cite{Schuett10} is in qualitative agreement with previous calculations~\cite{qcgraphene,MHDgraphene} and our results here, which, however, yield the {\em exact} leading order in $\alpha$.} 

\subsection{Spin current}
Here we are interested in the spin current driven by a magnetic field gradient. The dc magnetization current density due to moving electronic spins is given by
\begin{equation}
\vec{j}_s 
= \mu_s \sum_{\nu \s \l} \int \frac{\mathrm{d} ^2{k}}{(2 \pi)^2} \s \vec{v}_{\l \vec{k}}  \langle \gamma_{\nu \s \l \vec{k}}^{\dagger}  \gamma_{\nu \s \l \vec{k}} \rangle \equiv 
\mu_s \sum_{\nu \s \l} \int \frac{\mathrm{d} ^2{k}}{(2 \pi)^2} \s \vec{v}_{\l \vec{k}}  f_{\nu \s \l \vec{k}},
\label{spincurrent}
\end{equation}
where $\mu_s\approx \mu_B$ is the magnetic moment of the electrons, $\vec{v}_{\l \vec{k}}= \lambda \frac{\vec{k}}{k}v_F$ is the group velocity,
and $f_{\nu \s \l \vec{k}}$ is the occupation number of the electronic quasiparticles. For a discussion of coherent contributions to the current at finite frequency, see Ref.~\cite{ssqhe}.
In linear response to a gradient in an external Zeeman field $\vec\nabla B$ (perpendicular to the graphene plane, see Fig.~\ref{fig:spinconductivity}) we write for the latter
\bea
f_{\nu \s \l \vec{k}} = f^0_{\lambda k}+ \td g_{\sigma \lambda \vec{k}} f^0_{\lambda k}(1-f^0_{\lambda {k}}) \mu_s |\vec\nabla B|,
\eea
where $
f^0_{\lambda k} = 1/(e^{[\hbar v_F \lambda k - \mu)/T} +1]$
is the equilibrium distribution function, and  
\bea
\td g_{\sigma \lambda \vec{k}} =  \vec{v}_{\l \vec k} \cdot \vec{e} \,g_{\sigma \lambda}(k), \quad {\rm with}\quad \vec e =\frac{\vec\nabla B}{|\vec\nabla B|}.
\eea
Since $f$ and $g$ do not depend on the valley index $\n$, it is suppressed here and below.
Finally, the spin conductivity is defined as
\begin{eqnarray}
\label{defsigmas}
\sigma_s =\frac{\langle j_s\rangle}{|\vec \nabla B|} &=& 2  \sum_{\sigma \lambda} \int \frac{\mathrm{d}^2{k}}{(2 \pi)^2} \s \, \vec e\cdot \vec{v}_{\l \vec k}  \,\td g_{\sigma \lambda \vec{k}} f^0_{\lambda k}(1-f^0_{\lambda {k}}).
\end{eqnarray}

\subsection{Boltzmann equation}
\begin{figure}[!ht]
\begin{center}
 \includegraphics[width=8.5cm]{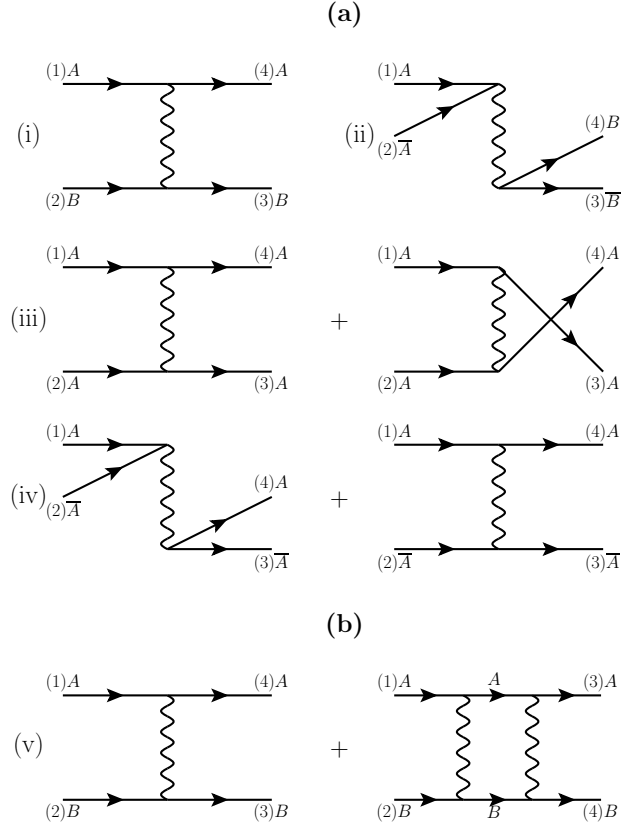}
 \caption{(a) The four classes of lowest order tree level Coulomb scattering processes, which enter the collision integral at leading order in $\alpha$. The labels $A,B$ stand for quantum numbers $(\td \s,\td \a,\lambda)$, $\overline{A}$ being the "antiparticle" of $A$ with all indices opposite. Apart from a factor which depends on the external momenta $\vec{k}_{1,...4}$, the matrix elements are proportional to $\l_A\l_B$, but independent of the indices $\td \a,\td \s$. Under the exchange of indices $(\td \s,\l)\to (\l,\td\s)$ the two interfering diagrams (iii) and (iv) are multiplied by the same factor $\pm 1$, so that the square of their sum is invariant at the level of the Born approximation. At the Dirac point the occupation factors $f^0_{\l,\td k}=f^0_{\td k}$ are also invariant. Hence, the collision operator acquires an accidental $SU(8)$ symmetry (if valley degrees of freedom are included) to leading order in the coupling $\a$.
(b) This does not hold anymore to order $O(\alpha^3)$ in the scattering rate. This is illustrated by the interfering diagrams (v), whose sum squared is not invariant under $(\td \s,\l)\to (\l,\td\s)$ if $\l_A\l_B\neq \td \s_A \td \s_B$.}
  \label{fig:matrixelements}
\end{center}
\end{figure}

The deviation function $\td g$ has to be found by solving a kinetic equation. For weak enough interactions, quasiparticles remain well defined, and the problem reduces to solving a semiclassical Boltzmann equation for $f_{\s\l\vec{k}}$ in the presence of a static driving field:
\bea
\frac{1}{\mu_s |\vec{\nabla}B|}  
\left[ \frac{\mathrm{d}f_{\sigma \lambda \vec{k}}}{\mathrm{d}t}\right]_{\mathrm{drive}} =  \frac{\sigma }{T} \vec e \cdot \vec{v}_{\l \vec k} f^0_{\lambda k} (1-f^0_{\lambda k}) \equiv  
 D_{\s\l \vec k}
\nn\\
\stackrel{!}{=}
-\frac{1}{\mu_s |\vec{\nabla}B|} \left[ \frac{\mathrm{d}f_{\sigma \lambda \vec{k}}}{\mathrm{d}t}\right]_{\mathrm{coll}} = \sum_{\s'\l'}\int\frac{d^2k'}{(2\pi)^2}C_{\s\l,\s'\l'}(\vec{k},\vec{k}')\td g_{\s'\l'\vec{k}'} \equiv [\hat C \td g]_{\s\l \vec k}.
\label{field2}
\eea
Here we have defined the driving term $D_{\s\l\vec{k}}$, with which the Boltzmann equation can be written compactly as $D=\hat C \td g$.
The kernel of the collision integral, $C_{\s\l,\s'\l'}(\vec{k},\vec{k}')$ defines the linear integral operator $\hat C$, which has been described in detail in previous works concerned with thermal and electrical transport~\cite{qcgraphene, MHDgraphene}. The form of $\hat C$ can be easily obtained from those references, taking into account that in the present case the dependence of $\td g$ on $\s$ requires to retain the $\s,\s'$-dependence of $\hat C$.
The two main ingredients of the kernel $\hat C$ are the occupation factors of in- and out scattering particles and the transition matrix elements, which enter the Fermi Golden rule for transition rates. At the level of the Born approximation, i.e., at the lowest order in the coupling $\alpha$, the relevant transitions are depicted in Fig.~\ref{fig:matrixelements}.

As we assume weak interactions, we can neglect the potential long ranged contributions to the collision integral due to the exchange of collective modes~\cite{Catelani}. The latter are known to contribute to thermal transport, but they are expected to affect spin and electrical conductivity at best at subleading order in the interactions. 

\subsection{Variational principle for the spin conductivity}
From Eqs.~(\ref{defsigmas},\ref{field2}) one sees that the solution of the Boltzmann equation satisfies $\sigma_s = 2 T \langle \td g,D \rangle$. This allows one to derive a convenient variational characterization of the spin conductivity (cf.~\cite{ziman,yaffe})
\begin{equation}
\label{maxprinciple}
\sigma_s = 2T\mu_s^2 \max_{\td g} \left [ \frac{\langle \td g, D \rangle ^2}{\langle \td g, \hat{C} \td g \rangle} \right],\quad {\rm with} \quad \langle {\td g},{\td h} \rangle \equiv \sum_{\s\l}\int \frac{d^2k}{(2\pi)^2}  {\td g}_{\s\l\vec k} {\td h}_{\s\l\vec k}.
\end{equation}
Using the positivity and Hermiticity of the collision kernel, one can show that the solution of the Boltzmann equation maximizes the ratio in (\ref{maxprinciple}).

The only dimensionless parameters of the problem at hand are $\alpha$ and the ratio $\mu/T$. Thus, apart from the quantum unit of $\mu_s^2/\hbar$, the spin conductivity is solely a function of these two parameters, $\s_s=\mu_s^2/\hbar\, \psi(\mu/T;\a)$. In the remainder of this paper we compute the scaling function $\psi$, its minimum at $\mu/T = 0$ and its asymptotic behavior at high chemical potential, $\mu/T \gg 1$.




\section{Solution of the Boltzmann equation at the Dirac point} 
Graphene has four species of massless Dirac quasiparticles, $(\nu,\s)=(\pm 1,\pm 1)$, each one coming as particle (conduction band) and hole (valence band) excitations ($\l =\pm 1$). 
At the Dirac point, these 8 types of excitations are all equally populated by a finite temperature. The corresponding $SU(8)$ symmetry of the non-interacting Hamiltonian is reduced by the  Coulomb interactions to a $SU(4)\times U(1)$ symmetry. However, as we will see below, at weak coupling this reduction of symmetry only manifests itself beyond the leading order interaction effects in the transport properties.

Instead of working with valence band electrons with quantum numbers $(\sigma,\lambda,\vec{k},\epsilon_k=\lambda \hbar v_F |k|)$ and deviation function $\td g_{\s\lambda \vec{k}}$, it is convenient to introduce hole excitations of positive energy with quantum numbers $(\tilde \sigma=\lambda \sigma,\lambda,\tilde{\vec{k}}=\lambda\vec{k},\tilde\epsilon_k=\lambda \epsilon_k)$ and group velocity $\vec{v}=d\epsilon_k/d{\vec k}= d\tilde\epsilon_k/d{\tilde{\vec k}}=\tilde{\vec{k}}/\tilde k$. \footnote{The valley index $\n$ changes under such a transformation, too, $\td\n =  \lambda \n$.} 
Further we introduce the deviation functions $\td h_{\td\s\l\vec{\td k}}= \lambda  \td g_{\s\l\vec k}$, and $h_{\td\s\l}(\td k) =\l g_{\s\l}(k)$. Positive $\td h_{\td\s\l \td{\vec  k}}$ corresponds to an excess of positive energy excitations as compared to equilibrium. 
A corresponding conjugation of the collision operator and the driving term lead to the Boltzmann equation in the form $\td D = \td{ C}\td h$, where
\bea
\td C_{\td \s  \l; \td \s'  \l' }(\td{\vec k},\td{\vec k'})  =\l C_{\s \l;\s' \l'}(\vec{k},\vec{k'}) \l' \nn\\ 
\td D_{\td \s \l \td{\vec k}}=\l D_{\s \l \vec{k}}.
\eea
In the case of spin and electrical conductivity (where we define $\vec e$ as  $\vec E/|\vec E|$) the driving terms are  
\bea
\td D^{s}_{\td \s \l \td{\vec k}} = \td \s \frac{\vec e \cdot \vec v_{\td {\vec k}}}{T}f^0_{\tilde k}(1-f^0_{\tilde k}), \quad \quad \td D^{e}_{\td \s \l \td{\vec k}} = \l \frac{\vec e \cdot \vec v_{\td {\vec k}}}{T}f^0_{\tilde k}(1-f^0_{\tilde k}),
\eea
respectively. 
Note that at the Dirac point, in the absence of a background magnetic field, 
\bea
\label{f0}
f^0_{\td k}=\frac{1}{\exp(\hbar v_F\td k/T)+1}
\eea
only depends on $\td k$, since $\mu=B=0$.
Thus the driving terms $\td D^{s,e}_{\td\s\l \td{\vec  k}}$ transform into each other under the exchange of 
indices $(\td \s,\l)\to (\l,\td \s)$. 
Formally, 
\bea
\label{drivetrafo}
\td D^{s} = \hat T \td D^{e},
\eea
where we define the operator $\hat T$ with $T_{\td \s \l,\td \s' \l'}=\delta_{\td \s\l'}\delta_{\l\td \s'}$.

For electrical transport, it has been shown in Ref.~(\cite{kashuba, qcgraphene}) that at the Dirac point the solution of the Boltzmann equation takes the form 
\bea
g^{e}_{\s \l}(k) = g^{De}(k)\quad {\rightarrow}\quad h^{e}_{\tilde\sigma\lambda}(\tilde k)=  \lambda  g^{De}(\tilde k).
\eea

From the symmetries of the Hamiltonian and the driving term, it follows similarly that in the case of magnetic driving the solution must take the form 
\bea
g^{s}_{\s \l}(k) = \s g^{Ds}(k) \quad {\rightarrow}\quad h^{s}_{\tilde\sigma\lambda}(\tilde k)=  \td \s  g^{Ds}(\tilde k).
\eea
In both cases the deviation from equilibrium, $h^{e,s}$ only depends on the sign of the coupling of the quasiparticles to the external driving field, but is otherwise independent of the quantum numbers $\lambda,\tilde \sigma$, and $\nu$. 

A detailed calculation of the spin conductivity shows that within the Born approximation the two solutions at the Dirac point are actually identical, $g^{Ds}(k)=g^{De}(k)$.  As a consequence, the dimensionless charge and spin conductivities are the same within the Born approximation (i.e., up to relative corrections of order $O(\alpha)$), 
\bea
\label{sigmaequal}
\frac {\s_s^{\rm Born}(\m=0)}{\m_s^2/\hbar} = \frac{\s_e^{\rm Born}(\m=0)}{e^2/\hbar}.
\eea
This result may come as a surprise since the Coulomb interaction, not being invariant under $\hat T$, breaks the $SU(8)$ symmetry of the non-interacting Hamiltonian and hence should  lead to different values of the transport coefficients. 
Indeed, we will show below that $\sigma^e$ and $\sigma^s$ differ at higher orders in the coupling strength $\alpha$. 

The reason for the accidental equality  at the level of the Born approximation can be understood as follows. Upon exchanging simultaneously 
all quantum numbers $\lambda$ and $\tilde \sigma$ of incoming and outgoing particles,
the transition amplitudes entering the collision integral change at most their sign. 
Applying Fermi's golden rule at the lowest order of perturbation theory in the collision integral, one finds that only matrix elements with identical transformation behavior  interfere with each other, cf.~Fig.~\ref{fig:matrixelements}. The common sign thus disappears upon squaring, and the resulting collision operator turns out to be invariant under the exchange of $\lambda$ and $\tilde \sigma$,
\bea
\label{symColl}
\hat T {\td { C}}^{(\rm Born)} \hat T^{-1} = \td { C}^{(\rm Born)}.
\eea
With Eq.~(\ref{drivetrafo}) this immediately entails the relation $\td h^s = \hat T \td h^e$ between the solutions, and thus the claimed equality of $g^{Ds}$ and $g^{De}$, and in particular Eq.~(\ref{sigmaequal}). 

It is quite clear that to higher order in $\alpha$ transition matrix elements with different transformation behavior under $\hat T$ do interfere, which lifts the invariance of the collision operator found to lowest order. This is illustrated in Fig.~\ref{fig:matrixelements}~(b), which shows two interfering transition amplitudes whose contribution to the collision rate is sensitive to the exchange of $\l$ and $\td \s$ at order $O(\a^3)$. 

A quantitative calculation of such higher order effects is very involved due to the large number of diagrams to be included in the collision operator. This is beyond our scope here. 
However, it may be interesting (and much simpler) to calculate the difference of spin and charge conductivities in examples of strongly coupled critical theories, which can be exactly solved by the AdS-CFT correspondence~\cite{AdSCFTviscosity, Herzog, nernst, AdsCFT}. An experimental measurement of both conductivities and in particular their difference, might provide an interesting test for the presence of strong interactions. On one hand such an experiment could quantitatively compare the two conductivities, which are both of the order of their natural quantum units. On the other hand it could potentially disentangle interaction effects from disorder effects, which so far dominate the existing experiments~\cite{vanWees09}.



\section{Spin conductivity as a function of chemical potential}

In this section we work with units in which $k_B=\hbar=v_F=1$, reinserting them only in final results.

\subsection{Separation of relaxation time scales in two dimensions}
Scattering problems in two dimensions have the interesting property that small angle forward scattering is strongly enhanced, without destroying immediately the Fermi liquid, unlike in one dimension~\cite{Maslov,qcgraphene}. This phenomenon introduces a separation of time scales between the relaxation which establishes equilibrium among quasiparticles with equal direction of motion, and the relaxation to global equilibrium. This feature was exploited in previous work, which mostly focussed on the critical window $\mu\lesssim T$, see Refs.~\cite{qcgraphene, MHDgraphene,landau}. Since, the respective timescales are parametrically different at weak coupling, one can obtain an approximate, but asymptotically (in small $\alpha$) exact solution  of the Boltzman equation by projecting it onto the slow relaxation modes. At the Dirac point this approximation is parametrically controlled by the logarithm of the weak coupling strength. As we will show below, the approximation becomes even much better at finite chemical potential for weakly screened Coulomb interactions, which strongly enhances small angle scattering and the tendency to establish partial equilibrium among particles with equal group velocity. 
 
\begin{figure}[!ht]
\begin{center}
 \includegraphics[height=6.5cm]{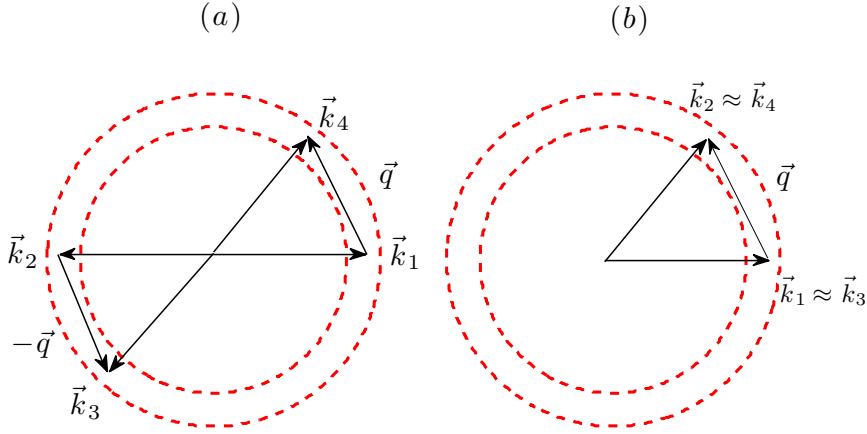}
  \caption{The dominant scattering channels in the degenerate Fermi liquid regime ($\mu\gg T$) consist in (a) head-on collisions  with $\vec k_1+\vec k_2 \approx \vec k_3-\vec k_4\approx 0$, and (b) scattering events with nearly identical pairs of incoming and outgoing momenta $(\vec k_1,\vec k_2) \approx (\vec k_3,\vec k_4)$ or $(\vec k_4,\vec k_3)$  [not shown]. For generic distribution functions $\td g$ small angle scattering $q\ll \mu$ dominates the scattering cross-section.
The relaxation of the slow mode, $\td g\propto \s \vec k/k $, proceeds essentially via the two shown  channels, involving scattering between opposite spins $\s_1=-\s_2$.}
\label {fig:smallanglescattering}
\end{center}
\end{figure}

The scale of fast relaxation rates of generic deviation profiles $\td g_{\s\l}(k)$ can be estimated as
\bea
\label{relrate}
\hbar \tau^{-1}_{\rm fast}(\td g)\sim \frac{\langle \td g|\hat C| \td g \rangle}{\langle \td g|f^0_{\l k} (1-f^0_{\l k})|\td g \rangle} \sim \frac{\langle \td g|\hat C| \td g \rangle}{T{\rm max}(T,\mu)} ,
\eea
where the denominator ${\langle \td g|f^0_{\l k} (1-f^0_{\l k})|\td g \rangle} \equiv \sum_{\s\l} \int \frac{d^2 k}{(2\pi)^2} \td g_{\s\l\vec k} f^0_{\l k} (1-f^0_{\l k}) \td g_{\s\l\vec k}$ normalizes the collision rate. 
The collisions at $\mu \lesssim T$ are dominated by forward scattering events. Their logarithmically divergent cross section is only cut-off by life-time broadening of the quasiparticles or higher order corrections in the interactions.~\cite{qcgraphene}  
This leads to a scattering rate
\bea
\label{fastDirac}
\hbar \tau^{-1}_{\rm fast}(\mu\lesssim T) \sim (V_0T)^2 T\log ([V_0T]^2)
\eea
where $V_0=V(q\sim T)$ is a typical interaction strength at thermal momentum transfers~\cite{qcgraphene} ($V_0 T\sim \alpha$ for Coulomb interactions, while for short range interactions $V(q)= V_0$, independently of $q$).
In the gated case, $\mu\gg T$, the fast relaxation is still dominated by forward scattering. Taking into account kinematic constraints, it can be estimated as an integral  
\bea
\hbar \tau^{-1}_{\rm fast}(\mu\gg T) \sim \frac{T^2}{\mu}\int_0^{2\pi} d\phi \frac{1}{|\sin(\phi)|} 
\left( 
\left[\mu V(q=2\mu \sin(\phi/2)\right]^2
+[\mu V_0]^2\right),
\eea
The first term refers to scattering processes as illustrated in Fig.~\ref{fig:smallanglescattering}, $\phi$ being the scattering angle between incoming and outgoing momenta, $\vec k_1$ and $\vec k_4$.
The second term is due to small angle scatterings which essentially preserve momenta, $\phi$ now being the angle between incoming momenta $\vec k_{1,2}$. 
The  factor $1/|\sin(\phi)|$ is crucial and originates from two-dimensional kinematics and phase space constraints. The logarithmic divergence at $\phi=0$, when all particles move parallel to each other, is only cut off by lifetime or screening effects, while the divergence at $\phi=\pi$ is regularized at angles $|\pi-\phi|\sim T/\mu$. For short range interactions the integral contributes a factor $(\mu V_0)^2$ 
and an additional logarithm $\log(\mu/\delta E)$, where $\delta E$ is the life time broadening of quasiparticles.
However, for weak Coulomb interactions and $T/\mu< \alpha$, the integral is strongly dominated by small momentum transfers $q\sim q_0\ll \mu$, leading to
\bea
\label{fastCb}
\hbar \tau^{-1}_{\rm fast, Cb}(\mu\gg T) \sim \frac{T^2}{\mu}\log(\a \mu/\delta E).
\eea

The strong forward scattering leads to a rapid equilibration of quasiparticles moving in the same direction. The remaining slow modes correspond to deviation functions $\td g$, where each subset of quasiparticles with the same $\vec v_{\vec k}$ is already in equilibrium. The only such  function $\td g^0_{\s\l}(\vec{k})$ which is also compatible with linear response (azimuthal dependence $\td g\sim \vec e\cdot \vec v$ and spin dependence $\td g\sim \s$) is 
\bea
\td g^0_{\s\l \vec{k}}= {\rm const.} \times \s \vec e \cdot\vec v_{\l \vec k}.
\label{slowmode}
\eea
Indeed, the quasi-equilibrium along fixed angles suppresses small angle scattering, the slow relaxation for the gated case being estimated as
\bea
\hbar \tau^{-1}_{\rm slow}(\mu\gg T) \sim \frac{T^2}{\mu}\int_0^{2\pi} d\phi \frac{\sin^2(\phi/2)}{|\sin(\phi)|} \left[\mu V(q=2\mu \sin(\phi/2)\right]^2,
\eea
where the extra factor $\sin^2(\phi/2)$ arises from the fact that quasiparticles traveling at similar angles are nearly equilibrated (see also Eqs.~(\ref{C002},\ref{C00general}) below). 
For short range interactions this rate scales as $(T^2/\mu) (V_0\mu)^2$, which is logarithmically smaller than the fast relaxation rate. For weak Coulomb interactions the ratio between the slow rate 
\bea
\hbar \tau^{-1}_{\rm slow, Cb}(\mu\gg T) \sim \alpha^2 \frac{T^2}{\mu}\log(1/\a),
\eea
and the fast rate (\ref{fastCb}) is even smaller, of order $O(\a^2)$, apart from logarithmic factors. 
This slow relaxation is dominated by head-on scatterings between opposite spins with small momentum transfer, see Fig.~\ref{fig:smallanglescattering}.
Likewise, close to the Dirac point, $\hbar \tau^{-1}_{\rm slow}(\mu\lesssim T)\sim (V_0T)^2 T$ is smaller than the fast mode (\ref{fastDirac}) by the logarithm which is absent in the slow relaxation.

\subsection{Spin conductivity from the slow mode approximation}
The ratio between slow and fast relaxation rates is a small parameter which allows us to obtain an excellent estimate of the spin conductivity by injecting the slow mode (\ref{slowmode}) into the maximum principle (\ref{maxprinciple}). 
Equivalently, we observe that the separation of time scales assures that the solution of the Boltzmann equation is proportional to the slow mode up to parametrically small corrections.
 
This principle has been used in Ref.~\cite{qcgraphene} to obtain the electrical conductivity at $\mu=0$ within a weak coupling approximation up to logarithmic corrections. As shown above the same numerical value applies to the spin conductivity, and thus
\bea
\sigma_s = \frac{0.121}{\alpha^2} \frac{\mu_s^2}{\hbar}.
\eea
The separation of time scales also allows for a parametrically controlled derivation of hydrodynamic equations from the semiclassical Boltzmann equation.~\cite{MHDgraphene,cyclotron}

\subsection{Degenerate Fermi liquid regime $\mu\gg T$ and 2d metals}

To leading order in $\mu/T$ the collisions are dominated by scattering of opposite spins, $(\vec k_1,\uparrow;\vec{k}_2,\downarrow)\to (\vec k_4,\uparrow;\vec{k}_3,\downarrow) $,
either with $\vec k_3\approx \vec k_1, \vec k_4\approx \vec k_2$ or with  $\vec k_1\approx -\vec k_2, \vec k_3\approx -\vec k_4$ (head-on collisions), see Fig.~\ref{fig:smallanglescattering}. These are the only momentum and energy conserving processes which are not strongly suppressed when the deviation function has relaxed to the slow mode, $\td g_{\s\l\vec{k}} \to  \td g^0_{\s\l\vec{k}}$.
In that case the collision matrix element in (\ref{maxprinciple}) simplifies to
\begin{eqnarray}
\langle \td g^0, \hat C \td g^0\rangle_{\mu\gg T}
\approx  4 \pi \int \frac{\mathrm{d} \vec{k}_1}{(2\pi)^2} \frac{\mathrm{d} \vec{k}_2}{(2\pi)^2}\frac{\mathrm{d} \vec{k}_3} {(2 \pi)^2} \delta(|\vec{k}_1|+|\vec{k}_2|-|\vec{k}_3|-|\vec{k}_4|) \nn \\
\quad \times F^0(k_1,k_2,k_3,k_4)   \left| T_{++++}(\vec{k}_1,\vec{k}_2,\vec{k}_3,\vec{k}_4) \right| ^2
 \left( \frac{\vec{k}_1}{k_1}-\frac{\vec{k}_2}{k_2}+\frac{\vec{k}_3}{k_3} - \frac{\vec{k}_4}{k_4} \right)^2
  \label{C002}
\end{eqnarray}
where $F^0$ is the product of occupation factors,
\begin{eqnarray}
F^0(k_1,k_2, k_3, k_4)&=& \frac{1}{e^{\b(k_1-\mu)}+1}\frac{1}{e^{\b(k_2-\mu)}+1}\frac{1}{e^{\b(\mu-k_3)}+1}\frac{1}{e^{\b(\mu-k_4)}+1} ,
\end{eqnarray}
while the matrix element $T_{++++}$ was defined in Eq.~(\ref{Ts}). The last term in (\ref{C002}) arises from the "transport factor" $[\td g^0(\vec{k}_1,\s)+\td g^0(\vec{k}_2,-\s) -\td g^0(\vec{k}_3,-\s)-\td g^0(\vec{k}_4,\s)]^2$. 
In the limit $\mu\gg T$, the integral~(\ref{C002}) can be evaluated analytically by observing that the last two factors in the integrand only depend on the scattering angle $\phi$ between $\vec k_1$ and $\vec k_4$, while the integrals over the other variables can be carried out analytically. The two dominant scattering channels shown in Fig.~\ref{fig:smallanglescattering} both contribute an equal amount to the matrix element, with the result
\begin{eqnarray}
\label{C00general}
\langle \td g^0, \hat C \td g^0\rangle_{\mu\gg T}
&\approx&  \frac{2 }{3\pi^2}T^3\int_{0}^{2 \pi} \mathrm{d} \phi \frac{\cos^4 \phi/2}{|\sin \phi |} \sin^2 \frac{\phi}{2} \left[\mu V\left(q=2 \mu \sin \frac{\phi}{2}\right)\right]^2.
\end{eqnarray}

For weakly screened Coulomb interactions, cf.~Eq.~(\ref{RPA}), this yields for $T/\mu\ll \a$
\begin{eqnarray}
\langle \td g^0, \hat C \td g^0\rangle_{\mu\gg T}^{\rm Cb}
&\approx&  \frac{8 \alpha^2}{3}T^3\int_{0}^{2 \pi} \mathrm{d} \phi \frac{\cos^4 \phi/2}{|\sin \phi |} \frac{ \sin^2 \phi/2}{({2}\sin \phi/2 + 4\alpha)^2}   \nn\\
&\approx &2\a^2 T^3 \left[\frac{2}{3}\ln\left(\frac{1}{2\alpha}\right)-1 +O(\alpha)\right].
\end{eqnarray}

The matrix element of the slow mode with the driving term evaluates simply to
\bea
\langle \td g^0,D\rangle = \frac{T}{\pi}\ln\left[2 \cosh(\mu/2T)\right]\approx \frac{\mu}{2\pi}.
\eea
We thus find the maximum of the spin conductivity functional as
\bea
\label{aymptotics}
\s_s^{\rm Cb}(\mu\gg T)  =   2T \frac{\langle \td g^0,D\rangle ^2}{\langle \td g^0, \hat C \td g^0\rangle} \frac{\mu_s^2}{\hbar}
= \frac{\mu_s^2}{\hbar} \frac{(\mu/T)^2}{4\pi^2\a^2} \left[\frac{2}{3}\ln\left(\frac{1}{2\alpha}\right)-1\right]^{-1},
\eea 
up to small relative corrections of order $O(T/\mu)$. 
The corresponding result for short range interactions tends to the $\mu$-independent limit,
\bea
\s_s^{\rm SR}(\mu\gg T)  =  \frac{3}{8\pi^2 (V_0 T)^2}\, \frac{\mu_s^2}{\hbar}.
\eea 

We confirmed numerically, by extending the maximization of the variational  principle to a larger function space $\td g$, that $\td g^0$ indeed gives an excellent approximation to the actual maximum of (\ref{maxprinciple}).

\begin{figure}[!ht]
\begin{center}
 \includegraphics[height=6.5cm]{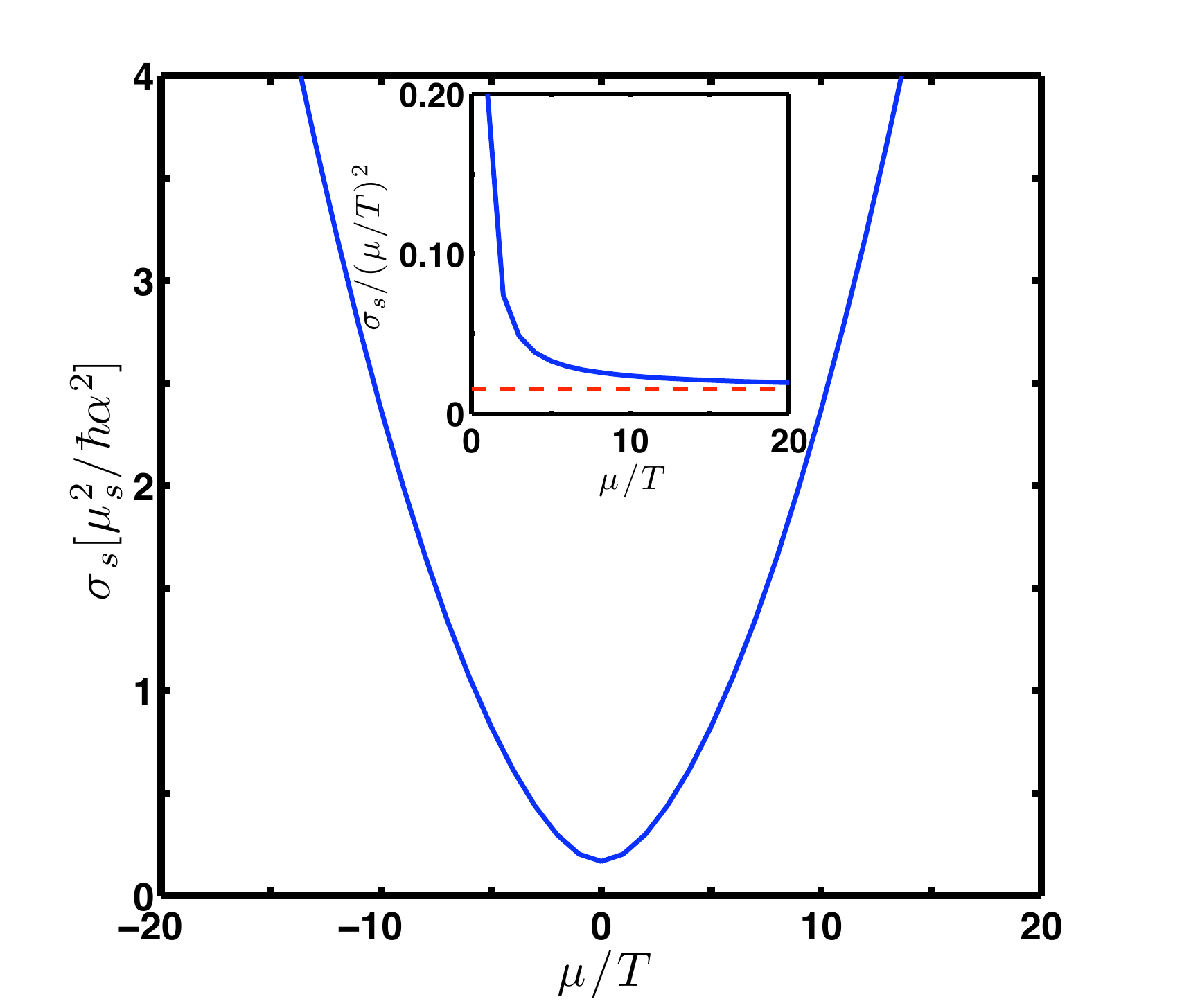}
 \includegraphics[height=6.5cm]{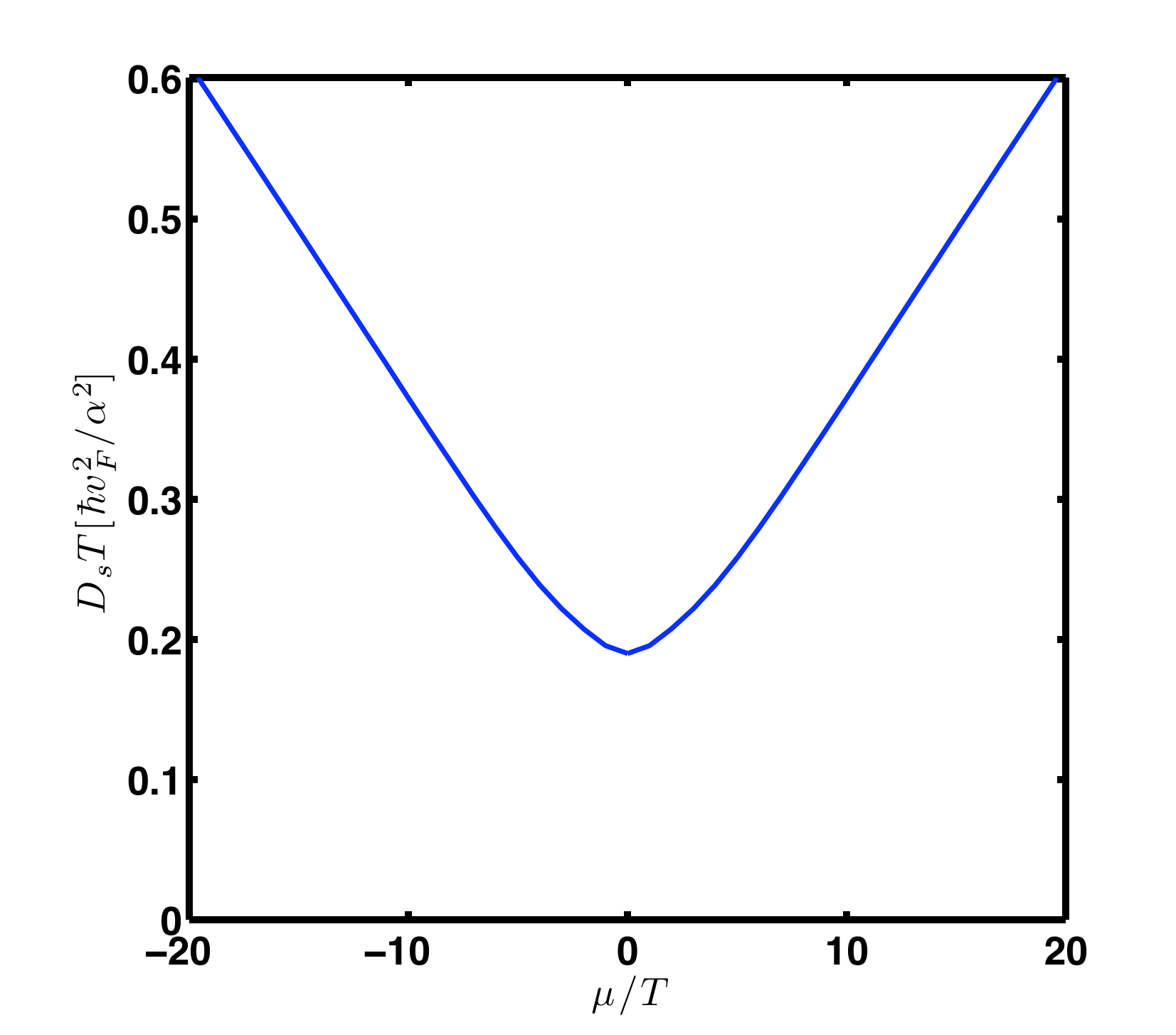}
  \caption{Spin conductivity $\s_s$ and spin diffusion constant $T D_s$ as a function of chemical potential. The diffusion constant behaves as $D_s\sim 1/T$ close to the Dirac point, and scales as $D_s\sim |\mu|/T^2$ in the degenerate Fermi regime $|\mu|\gg T$. The inset on the left shows the approach of $\s_s$ to the analytical prediction Eq.~(\ref{aymptotics}) (dashed line). To regularize small momentum transfers Thomas-Fermi screening has been incorporated assuming a coupling $\a = 0.01$. Note that the screening slightly increases the value of $\sigma_s(\m=0)$.}
  \label{fig:spinconductivityplot}
\end{center}
\end{figure}

For small and moderate values of $\mu/T$, $\langle \td g^0, \hat C \td g^0\rangle$ has to be be evaluated numerically. The resulting full scaling function for the spin conductivity is plotted in Fig.~\ref{fig:spinconductivityplot}. We also display the spin diffusion coefficient $D_s$, which is related to $\s_s$ by the Einstein relation 
$D_s= \s_s/\chi$
where $\chi= \frac{4}{\pi}\frac{\mu_s^2 T}{(\hbar v_F)^2}\ln[2\cosh(\mu/2T)]$
is the Pauli susceptibility. At the Dirac point $D_s\sim 1/T$, while in the degenerate Fermi liquid regime one finds $D_s\sim \mu/T^2$. 

The scaling $\sigma_s\sim \frac{\mu_s^2}{\hbar}  (\mu/T)^2$ is easy to understand with similar arguments as those given in the introduction for the conductivity at the Dirac point. The scaling is expected from a Drude-type estimate $\sigma_s \approx\mu_s^2 n \tau_{\rm inel}/m_{\rm eff}$,
where $m_{\rm eff}=p_F/v_F =\mu/v_F^2$ is the effective inertial mass and $n \sim (\mu/\hbar v_F)^2$ the density of carriers, while $\tau_{\rm inel}\sim \hbar \mu/T^2$ is the  inelastic {\em transport} scattering time of a 2d Fermi liquid. We stress that the transport time coincides with the slow relaxation time, which in two dimensions is parametrically smaller  than the collision time, which scales like $\tau_{\rm fast}$. The latter controls the quasiparticle life time, but not the transport. In contrast to the collision and life times, the transport time does not contain an extra logarithm of $\mu/T$, a fact which was missed in earlier studies of quasi-two dimensional metals~\cite{Wilkins}. 

In the case of screened Coulomb interactions the transport time is enhanced by a factor $\log(\alpha)$ which reflects the enhanced scattering at small momentum transfer. Apart from this logarithm the transport time is independent of the bare strength of the interactions $\a$.

\section{Discussion and Conclusion}
The minimal spin conductivity of the order of the quantum unit $\mu_s^2/h$  reflects the strong inelastic scattering among quasiparticles at finite temperatures close to the Dirac point. It is one of several interesting hallmarks of the strongly coupled electron-hole plasma, 
together with a minimal collision-dominated electrical conductivity and a minimal viscosity to entropy ratio at $\mu=0$. All these aspects are very similar to the phenomenology near quantum critical points, and arise here due to the marginal irrelevance of the Coulomb coupling.
However, the disorder free, collision-dominated spin conductivity is also well-defined away from the Dirac point, as well as in any clean metal without spin orbit coupling.
The analytical results we obtained for graphene in the degenerate Fermi liquid regime are general and immediately carry over with minimal adjustments to the case of 2d metals with simple Fermi surfaces. 

The simplest way to observe collision-limited spin diffusion consists presumably in preparing a local spin polarization by an external field, and monitoring its spreading (with diffusion constant $D_s$) after switching off the field. In reality this diffusion will be limited in space and time by a finite (even though rather long) spin flip time. 

Alternatively, the following Gedankenexperiment involves a spin current source to measure $\sigma_s$ in the analog of an electrical two-point measurement. Imagine to contact the sample by the two opposite edges of a spin Hall conductor~\cite{Hirsch}, in which  a spin Hall voltage is induced. This spin-Hall source then injects oppositely polarized currents into the sample, but no net current. 
The resulting total spin current in the sample will be controlled by the collision-limited spin conductance of the sample (proportional to $\sigma_S$, the width and the inverse length of the sample) and the contact resistances. 
  
We hope that future experiments in suspended graphene with unscreened Coulomb interactions will explore these interesting facets of collision-dominated transport.

\section{Acknowledgment}
We thank S. Sachdev for attracting our interest to the problem of spin conductivity. We thank him and L. Fritz for useful discussions.

\end{document}